\documentclass[sigconf]{acmart}

\usepackage{graphicx}
\usepackage{url}
\newcommand{\gestop}[0]{{\sc Gestop}}

\copyrightyear{2021}
\acmYear{2021}
\setcopyright{rightsretained}
\acmDOI{10.1145/3430984.3430993}

\acmConference[CODS COMAD 2021]{8th ACM IKDD CODS and 26th COMAD}{January 2--4, 2021}{Bangalore, India}
\acmBooktitle{8th ACM IKDD CODS and 26th COMAD (CODS COMAD 2021), January 2--4, 2021, Bangalore, India}
\acmPrice{15.00}
\acmISBN{978-1-4503-8817-7/21/01}

\begin{document}

\title{{\gestop}: Customizable Gesture Control of Computer Systems}

\author{Sriram S K}
\email{sriramsk1999@gmail.com}
\affiliation{%
  \institution{PES University}
  \city{Bengaluru}
  \country{India}
}

\author{Nishant Sinha}
\email{nishant@offnote.co}
\affiliation{%
  \institution{OffNote Labs}
  \city{Bengaluru}
  \country{India}}

\begin{abstract}
 The established way of interfacing with most computer systems is a mouse and keyboard. 
 Hand gestures are an intuitive and effective touchless way to interact with computer systems. 
 However, hand gesture based systems have seen low adoption among end-users primarily due to numerous technical hurdles in detecting in-air gestures accurately.
 This paper presents {\gestop}, a framework developed to bridge this gap. The framework learns to detect gestures from demonstrations, is customizable by end-users and enables users to interact in real-time with computers having only RGB cameras, using gestures.

\end{abstract}



\maketitle

\section{Introduction}

Hand detection and gesture recognition has a broad range of potential applications, including in-car gestures, sign language recognition, virtual reality and so on. Through gestures, users can control or interact with devices without touching them. 
Although numerous gesture recognition prototypes and tutorials are available across the web, they handle only restricted set of gestures and lack proper architecture design and description, making it hard for end-users to use and build upon them.

In this paper, we present the architecture and implementation of our end-to-end, extensible system which allows user to control desktop in {\it real-time} using hand gestures. Our tool works across different hardware (CPU/GPU) and operating systems and relies on a medium resolution camera to detect gestures. The tool controls the desktop through hand gestures alone, replacing all mouse actions with gestures, and many keyboard shortcuts as well. Furthermore, the design is {\it modular} and {\it customizable}: we provide an easy-to-use configuration for remapping gestures and actions, adding new custom actions as well as new gestures.

We make use of two kinds of gestures in our application: \textbf{static} and \textbf{dynamic}. Static gestures are gestures where a single hand pose provides enough information to classify the gesture, such as the "Peace" sign.
On the other hand, {\it dynamic} gestures cannot be detected from a single pose alone, and require a \textit{sequence} of poses to be understood and classified. Examples include the gestures maintaining the pose of the hand while moving it ("Swipe Up"), or gestures which involve changing hand posture continuously ("Pinch"). 
By combining hand motion along with continuous pose change, we can create a large number of dynamic gestures.


The architecture of the application follows a modular design. It is separated into logical components, each performing a single task.
The \textbf{Gesture Receiver} receives keypoints from the image and passes it on to the \textbf{Gesture Recognizer}, which uses neural networks
to classify both static and dynamic gestures, and finally the \textbf{Gesture Executor}, which executes an action based on the detected gesture.

A key distinguishing aspect of our application is that it is {\it customizable} by the end-user. Other than the inbuilt mouse and keyboard functions,
it is possible to map gestures to arbitrary desktop actions, including a shell script. 
This allows for massive
flexibility in how the application can be used. Gestures can be mapped to launch other applications, setup
environments and so on. In addition, we have provided a way to add new gestures as well, allowing the user to extend it as much as required.

\section{Related Work}

{\bf Hand Gesture Recognition from Video.} 
The literature on gesture recognition from static images or video is vast.
The solutions vary based on whether (a) the cameras or sensors (single or multiple instances) provide RGB-only images vs depth data (RGB-D), (b) we detect hand keypoints (palm and finger joints) as an intermediate step or perform end-to-end detection directly from video, (c) detecting keypoints is the end-goal as opposed to 3D reconstruction of hands, (d) gestures are pre-segmented or must be segmented in real-time.
For more details, please refer to recent overview articles by Lepetit~\cite{lepetit20} and Ren et al.~\cite{ren2020survey}.
In our approach, we use a monocular RGB video stream, which is fed into a two-phase neural network architecture. 
The first phase detects hand keypoints from individual video frames (using the off-the-shelf MediaPipe~\cite{mpipe} tool) and generates a sequence of hand keypoints, which is used by the second phase to detect both static and dynamic gestures (using our neural models).
GestARLite~\cite{gestarlite} is a light, on-device framework for detecting gestures based on pointing fingers. 
Our solution is targeted towards desktop computers and can recognize much larger set of complex dynamic gestures and can be customized easily.

{\bf Gesture Recognition Platforms.}
Gesture recognition is useful for several applications: controlling virtual interfaces, gaming, embodied AR/VR environments, automotive human-machine interface~\cite{amara19}, home automation~\cite{ruserhome20}, education~\cite{streeter20}, retail business environments, consumer electronics control and more. 
Many applications use hand gestures because they enable highly expressive interaction.
Although the gesture recognition market size is rapidly growing, there are hardly any end-to-end open source gesture recognition platforms.
Exceptions include GRT~\cite{grt}, an open-source, C++ machine learning library designed for real-time gesture recognition, which provides building blocks for creating custom recognizers.
In contrast, we use neural layers as our building blocks and develop models using PyTorch~\cite{pytorch}.

Programming imperative multi-touch gesture recognizers involves dealing with low-level, event-driven programming model. 
Gesture Coder~\cite{gcoder} learns from multi-touch gesture examples provided by users and generates imperative recognition code automatically and invokes corresponding application actions.
Oney et al.~\cite{oney19} investigate declarative abstractions for simplifying programming of multi-touch gestures.
In contrast, we recognize gestures end-to-end using deep learning from video examples.

\section{Design Architecture}

We use an open source framework (\textit{MediaPipe}) \cite{lugaresi2019MediaPipe} to detect the hand 
keypoints in the images captured by the camera. 
The \textbf{MediaPipe} module reads data from the camera, processes it and generates keypoints which 
are then sent to the \textbf{Gesture Receiver} using ZeroMQ, a messaging queue. On receiving the keypoints, the 
\textbf{Gesture Receiver} then passes it to the \textbf{Gesture Recognizer}, which then processes the keypoints
into the encoded features, feeds them into the network which detects the output gesture. Finally, the 
\textbf{Gesture Receiver} sends the detected gesture to the \textbf{Gesture Executor}, which executes an action.

\begin{figure}[h!]
    \centering
    \includegraphics[scale=0.15]{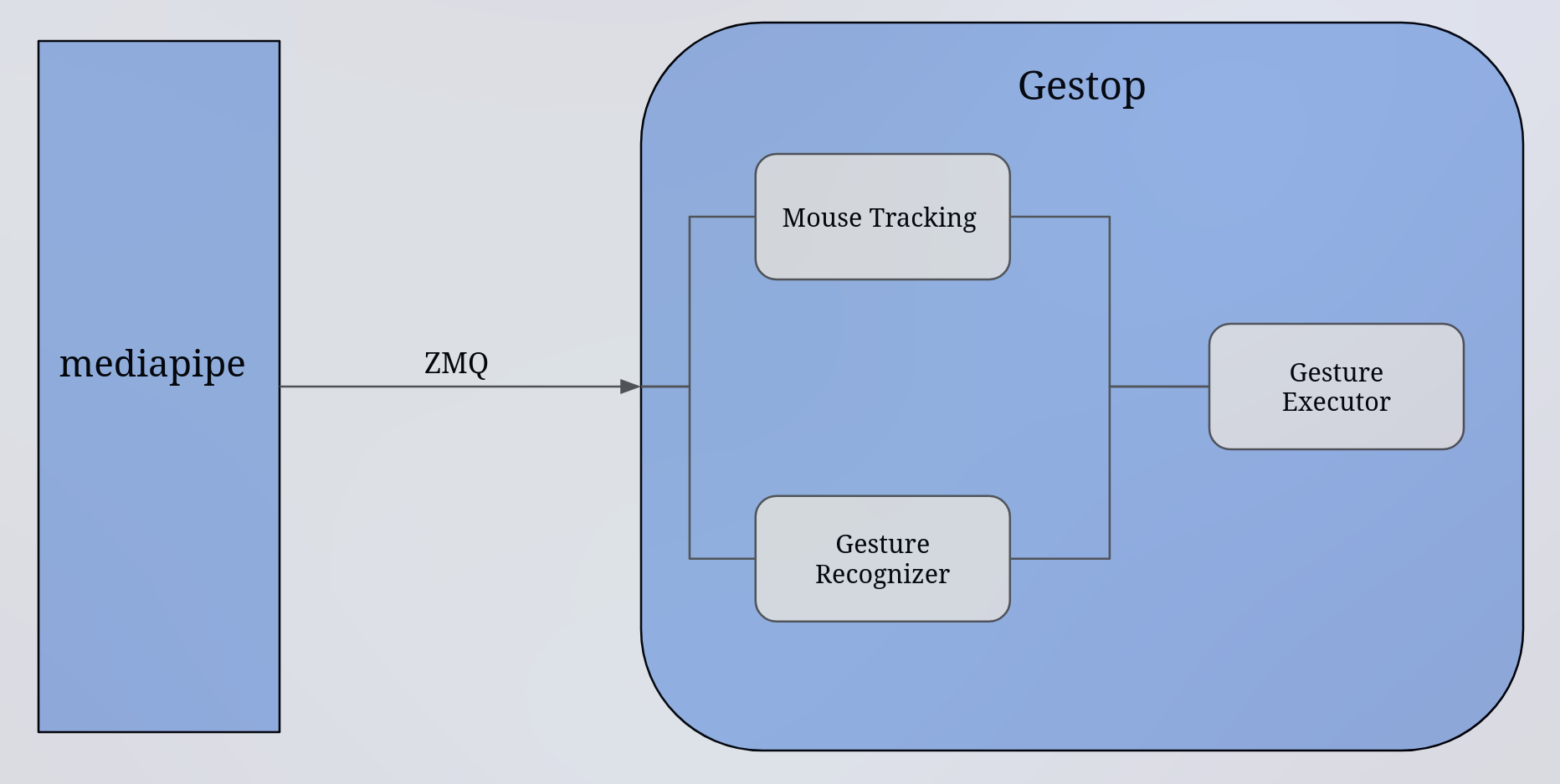}
    \caption{Overview of design architecture}
    \label{fig:design_arch}
\end{figure}

\textbf{MediaPipe}. The first component, which tracks the palms of the user and generates the hand landmarks or 
keypoints is built using \textbf{MediaPipe}, a cross-platform framework providing a variety of ML solutions. We
utilize MediaPipe's Hand Tracking \cite{zhang2020MediaPipe}, a high-fidelity hand and finger tracking solution 
which can infer 21 3D landmarks of a hand from just a single frame. The tracking is smooth and handled cases of 
self-occlusion (the hand covering itself) as well.  

\textbf{Gesture Receiver}. The Gesture Receiver is the heart of the application and acts as a controller for the 
other modules. It receives the keypoints from the MediaPipe module, and then passes it to the \textit{Gesture 
Recognizer} and the \textit{Mouse Tracker}. The output received (the name of a gesture) is then passed to the 
\textit{Gesture Executor}.

\textbf{Mouse Tracker}. The Mouse Tracker tracks the cursor on the screen as the hand moves. As convention, we use 
the tip of the index finger as the keypoint with which to track the mouse. As the index finger is moved, 
the motion is projected onto the screen and the cursor moves accordingly.

\textbf{Gesture Recognizer}. The Gesture Recognizer module classifies gestures given keypoints. We utilize two 
neural networks for the same, one to detect static gestures and the other for dynamic gestures. The details of 
their structure and training are elaborated upon in the subsequent sections.

\textbf{Gesture Executor}. The input to this module is the name of the gesture which has been 
recognized by the Gesture Recognizer. This module finds the action mapped to this gesture and then executes it.
We include a small set of predefined gestures and actions to cover common use cases. 

\section{Gesture Recognizer}
The inputs to the Gesture Recognizer are the 21 3D keypoints generated by MediaPipe. each corresponding to a point
on the hand. Each keypoints consists of three coordinates $(x,y,z)$. Thus the Gesture Recognizer receives a 63-D input vector. These input vectors are then transformed into the features expected by the neural networks as described below. 

\subsection{Static Gestures}
\begin{figure}[h!]
\setlength\tabcolsep{2pt}
\centering
\begin{tabular}{ccccc}
 \textbf{Eight} &
 \textbf{Four} &
 \textbf{Hitchhike} &
 \textbf{Seven} &
 \textbf{Spiderman} \\
 \includegraphics[scale=0.2]{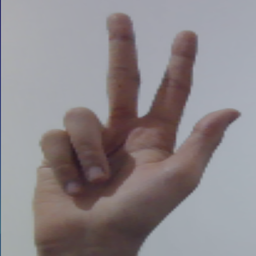} &
 \includegraphics[scale=0.2]{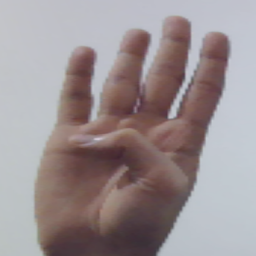} &
 \includegraphics[scale=0.2]{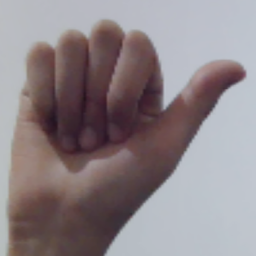} &
 \includegraphics[scale=0.2]{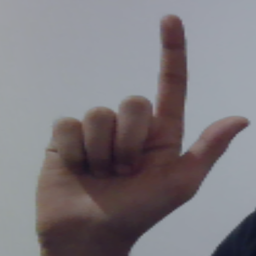} &
 \includegraphics[scale=0.2]{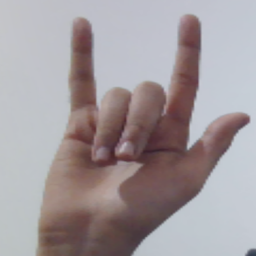}
\end{tabular}
\caption{Sample static gestures that are included with the application.}
\label{fig:gesture_list}
\end{figure}

Static gestures are gestures which can be described by a single hand pose. Some of the gestures detected by {\gestop} are shown in Fig.~\ref{fig:gesture_list}, along with the name which is used to refer to them.
The set of static gestures included in the tool can virtually replace the mouse for all uses. 

\textbf{Feature Computation}. The keypoints generated by MediaPipe are transformed and fed into the network. The vector is computed by calculating the \textit{relative hand vectors}: the vector differences between the input keypoints. 
These relative vectors encode hand pose information in a position invariant manner, i.e. the same gesture is detected, regardless of where the hand is in the webcam's field of vision.
For example, the first relative hand vector (from the base of the palm to the first joint of the thumb) can be computed by the following:

\begin{equation*}
    V01_x = V1_x - V0_x
\end{equation*}
\begin{equation*}
    V01_y = V1_y - V0_y
\end{equation*}
\begin{equation*}
    V01_z = V1_z - V0_z
\end{equation*}

\begin{figure}[h!]
    \centering
    \includegraphics[scale=0.25]{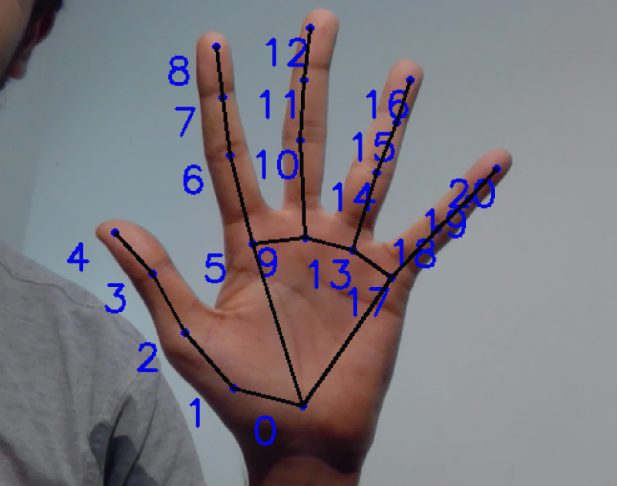}
    \caption{The labeled keypoints generated by MediaPipe}
    \label{fig:label_hand}
\end{figure}

Where $V0$ and $V1$ represent the 3D coordinates of the points labeled 0 and 1 in Fig. \ref{fig:label_hand} and $V01$ 
represents the \textit{relative} hand vector between them.
In summary, we have 16 relative hand vectors (4 for the thumb, 3 for the other fingers) and each hand vector 
consisting of $(x, y. z)$ coordinate giving us  a total of 48 coordinates. Finally, the \textit{handedness}, i.e., the hand with which the gesture is performed, is appended and this 49-D vector  is fed into 
the network (Sec.~\ref{sec:impl}).

\textbf{Dataset}. In training the network, we recorded our own data and created a small dataset.
The data was collected in the following manner: The name of the gestures is specified, the MediaPipe component is 
executed and keypoints were captured. These keypoints are simply written to a CSV file for later use along 
with the given gesture name. For each gesture, around 2000 samples were taken, which took only a couple of minutes.

While static gestures are simple to perform,
they are limited by virtue of the fact that they are only so many distinct poses one can perform with a hand.
Hence, dynamic gestures are required.

\subsection{Dynamic Gestures}

Dynamic Gestures are an extension of static gestures, and consist of a sequence of poses. These include gestures 
commonly used on touchscreen devices, such as "Swipe Up", "Pinch" etc.

\textbf{Dataset}. For training dynamic gestures, we make use of the SHREC \cite{de2017shrec} dataset. This dataset consists of 2800 
sequences across 14 gestures, including common ones like "Swipe Up", "Tap" as well as more complex gestures like "Swipe +".
It consists of variable length sequences performed by multiple people in 2 ways: either using the whole hand, or 
just the fingers. To capture more data, {\gestop} also includes a script to record dynamic gestures.

\textbf{Feature Computation}. To compute the feature vector, each frame of the input sequence is transformed into a vector consisting of the following:
\begin{itemize}
  \item The \textit{absolute} coordinates of the base of the palm i.e. $V0_x$ and $V0_y$. This was used because gestures like 
  "Swipe Up", "Swipe Right" etc. involve moving the hand.
  \item The \textit{timediff} coordinates of the base of the palm. This consisted of the change in position of that coordinate
  with respect to the previous timestep. This was empirically found to improve the performance of the network.
  \item Finally, similar to the static case, coordinates of the relative hand vectors to capture the \textit{pose} of the hand.
\end{itemize}

\section{Gesture Executor}

The Gesture Executor is the user-facing module. 
Its responsibility is to take in the recognized gesture, map it to the specified action and then 
execute it. The mapping of gestures to actions is stored in a human-readable JSON file, for easy modification. 
The format of the file is:

\begin{verbatim}
    {'gesture-name':['type','func-name']}
\end{verbatim}

Here, \textit{gesture-name} is the name of a gesture, \textit{type} is either \verb|sh| (shell) or \verb|py| 
(python), denoting the type of the action to be executed and \textit{func-name} is the name of the shell 
script/command to be executed if the type is \verb|sh|, or the name of a user defined function if the type if 
\verb|py|. To remap functionality, for example, if the user wishes to take a screenshot with \verb|Swipe +| instead of \verb|Grab|, the configuration would change from:

\begin{verbatim}
    "Grab" : ["py", "take_screenshot"],
    "Swipe +" : ["py", "no_func"],
\end{verbatim}

To,

\begin{verbatim}
    "Grab" : ["py", "no_func"],
    "Swipe +" : ["py", "take_screenshot"],
\end{verbatim}

{\bf Custom Actions.} To suit an end user's specific workflow, {\gestop} allows defining custom actions, e.g., a python function or a shell script, which are executed when the corresponding gesture is detected. For example, to execute a shell script \verb|script.sh| on the "Tap" gesture, the user may change the mappings in the configuration file to:

\begin{verbatim}
    {'Tap':['sh','./script.sh']}
\end{verbatim}

\subsection{New Gestures}

In our experience, the end user may want to add both new actions and new gestures. Hence, we have provided a method to add data for new gestures (static or dynamic).
For static gestures, the same script that was used to collect initial data to train the network is reused to add more gestures. The new gesture name is provided, and data is recorded and written to disk. For recording dynamic data, we provide a script: for each script run, a gesture name is provided and the corresponding gesture performed repeatedly.
Post-run, multiple sequences with gesture labels are stored on the disk. Data for a new gesture, 'Circle', was collected using this process over the course of 15-20 minutes, demonstrating its feasibility. 
After adding new gesture data, the network is retrained and the application is now able to detect the user's custom gestures as well.

\begin{figure}
    \centering
    \includegraphics[scale=0.25]{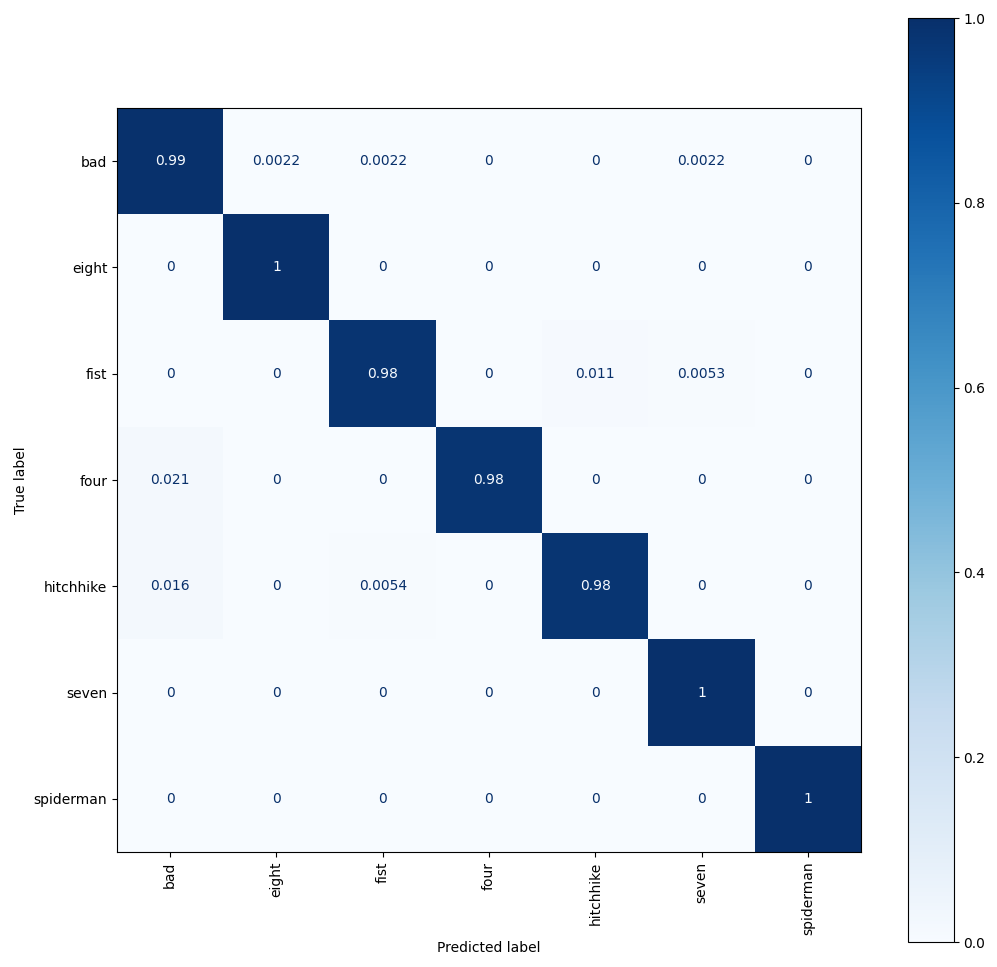}
    \caption{Confusion Matrix for static gestures}
    \label{fig:static_conf_mat}
\end{figure}

\section{Implementation and Results}
\label{sec:impl}
We use the pytorch-lightning 
\cite{falcon2019pytorch} framework to build our neural network classifiers. The implementation is open-sourced~\cite{gestop} with detailed documentation for installation and usage, along with a demo video showcasing {\gestop}'s capabilities. 

\textbf{Static Gestures}. To detect static gestures, we utilize a feed forward neural network classifier with 2 linear 
layers, which takes in a feature vector and classifies it as one of the available gestures. The network was trained for
around 50 epochs and the confusion matrix of the trained network can be seen in Fig. \ref{fig:static_conf_mat}. 

The set of static gestures relevant to an application are much smaller than all possible static gestures.
Moreover, it is infeasible to record all {\it unwanted} gestures to help our classifier discriminate accurately.
We handle this {\it data imbalance} as follows.
Besides the set of relevant gestures, we introduce a {\it none} gesture, which is selected if no relevant gesture is detected.
To train our classifier, we capture a variety of unrelated static gestures and label them as {\it none}.
While this improves classifier performance, we still see many false positives.
To solve this problem, we manually calibrate the {\it softmax} output of the classifier by scaling the score of the {\it none} gesture by a constant $k$ ($k = 2$ worked well for our experiments).

These optimizations allow our static classifier to achieve high detection accuracy for multiple users and different lighting conditions.
We achieve good performance (Fig.~\ref{fig:static_conf_mat}), with a validation accuracy of 99.12\%, and this translates to test time as well. Hand gestures are detected with no noticeable latency.

\begin{figure}
    \centering
    \includegraphics[scale=0.2]{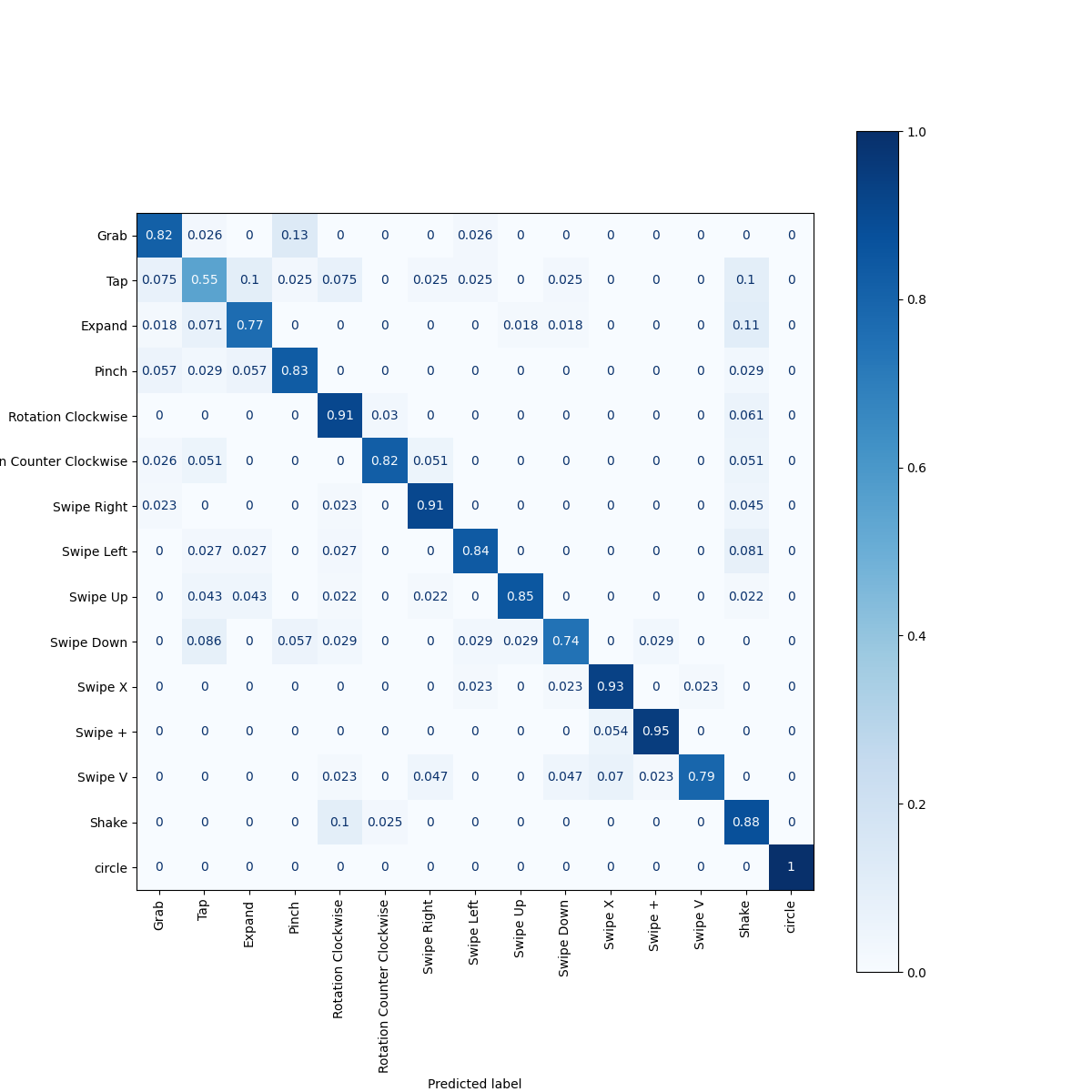}
    \caption{Confusion Matrix for dynamic gestures}
    \label{fig:dynamic_conf_mat}
\end{figure}

\textbf{Dynamic Gestures}. To detect dynamic gestures, we use a recurrent neural network, which consists of a linear layer, to encode the incoming features, connected to a bidirectional GRU. A key issue when detecting dynamic gestures is that of computing the start and end of a gesture precisely. We circumvent this issue by making use of a {\it signal} key to signify the start and end of the gesture; we utilize the \verb|Ctrl| key in {\gestop}.
This enables handling varying length gestures as well as reduces the number of misclassifications.
In addition to the gestures provided by SHREC, an additional gesture, 'Circle', was also added using the aforementioned methods. Despite being a complex gesture, the network was able to detect the gesture accurately during testing, leading us to believe that the network can successfully generalize to other gestures as well.  
The confusion matrix for the various gestures is shown in Fig. \ref{fig:dynamic_conf_mat}.
We observe that dynamic gestures have lower performance
than static gestures, with an average accuracy of around 85\%. 
Dynamic gestures are inherently concerned with 
two factors, the pose of the hand and the displacement of the hand over time. As can be seen from the confusion matrix, gestures which involve displacement of the hand (i.e. the "Swipes") are detected well, whereas those concerned with the orientation of the hand such as "Tap" have a relatively lower accuracy. 
When testing our models trained on SHREC dataset, we also observed a {\it domain mismatch} problem.
The data from SHREC was recorded using an Intel RealSense depth camera, whereas the incoming stream during testing is from an RGB camera, causing loss in accuracy during testing.
In our ongoing work, we plan to address both these issues with improved feature computation and training on larger, diverse datasets.

Compared to existing systems like GestARLite~\cite{gestarlite}, {\gestop} doesn't require a headset or additional hardware to operate. GRT-\cite{grt}, a gesture recognition toolkit in C++, provides building blocks for users to build a gesture recognition pipeline, whereas {\gestop} provides a complete pipeline and a simple interface for end users to customize.

\section{Conclusion}

In this paper, we present {\gestop}, a novel framework for controlling the desktop through hand gestures, 
which may be customized to preferences of the end-user. In addition to providing a fully functional replacement for the mouse, our framework is easy to extend by adding new custom gestures and actions, allowing the user to use gestures for many more desktop use cases.
We aim to improve {\gestop} further by improving the detection accuracy, make incremental training for new gestures efficient, detecting gesture start/end and other subtle user intents, and conduct user studies to measure the usability of the tool.

\begin{acks}

We would like to thank Vikram Gupta for useful discussions.

\end{acks}

\bibliographystyle{ACM-Reference-Format}
\bibliography{gesref}


\begin{thebibliography}{16}


\ifx \showCODEN    \undefined \def \showCODEN     #1{\unskip}     \fi
\ifx \showDOI      \undefined \def \showDOI       #1{#1}\fi
\ifx \showISBNx    \undefined \def \showISBNx     #1{\unskip}     \fi
\ifx \showISBNxiii \undefined \def \showISBNxiii  #1{\unskip}     \fi
\ifx \showISSN     \undefined \def \showISSN      #1{\unskip}     \fi
\ifx \showLCCN     \undefined \def \showLCCN      #1{\unskip}     \fi
\ifx \shownote     \undefined \def \shownote      #1{#1}          \fi
\ifx \showarticletitle \undefined \def \showarticletitle #1{#1}   \fi
\ifx \showURL      \undefined \def \showURL       {\relax}        \fi
\providecommand\bibfield[2]{#2}
\providecommand\bibinfo[2]{#2}
\providecommand\natexlab[1]{#1}
\providecommand\showeprint[2][]{arXiv:#2}

\bibitem[\protect\citeauthoryear{??}{ges}{2020}]%
        {gestop}
 \bibinfo{year}{2020}\natexlab{}.
\newblock \bibinfo{booktitle}{\emph{Gestop}}.
\newblock
\newblock
\shownote{\url{https://github.com/sriramsk1999/gestop}.}


\bibitem[\protect\citeauthoryear{??}{grt}{2020}]%
        {grt}
 \bibinfo{year}{2020}\natexlab{}.
\newblock \bibinfo{booktitle}{\emph{Gesture Recognition Toolkit}}.
\newblock
\newblock
\shownote{\url{https://github.com/nickgillian/grt}.}


\bibitem[\protect\citeauthoryear{??}{mpi}{2020}]%
        {mpipe}
 \bibinfo{year}{2020}\natexlab{}.
\newblock \bibinfo{booktitle}{\emph{{M}edia{P}ipe: Cross-platform ML solutions
  made simple}}.
\newblock
\newblock
\shownote{\url{https://google.github.io/mediapipe/}.}


\bibitem[\protect\citeauthoryear{Amara}{Amara}{2019}]%
        {amara19}
\bibfield{author}{\bibinfo{person}{Hassene~Ben Amara}.}
  \bibinfo{year}{2019}\natexlab{}.
\newblock \showarticletitle{End-to-End Multiview Gesture Recognition for
  Autonomous Car Parking System}.
\newblock


\bibitem[\protect\citeauthoryear{De~Smedt, Wannous, Vandeborre, Guerry,
  Le~Saux, and Filliat}{De~Smedt et~al\mbox{.}}{2017}]%
        {de2017shrec}
\bibfield{author}{\bibinfo{person}{Quentin De~Smedt}, \bibinfo{person}{Hazem
  Wannous}, \bibinfo{person}{Jean-Philippe Vandeborre}, \bibinfo{person}{Joris
  Guerry}, \bibinfo{person}{Bertrand Le~Saux}, {and} \bibinfo{person}{David
  Filliat}.} \bibinfo{year}{2017}\natexlab{}.
\newblock \showarticletitle{Shrec'17 track: 3d hand gesture recognition using a
  depth and skeletal dataset}.
\newblock


\bibitem[\protect\citeauthoryear{Falcon et~al\mbox{.}}{Falcon
  et~al\mbox{.}}{2019}]%
        {falcon2019pytorch}
\bibfield{author}{\bibinfo{person}{WEA Falcon} {et~al\mbox{.}}}
  \bibinfo{year}{2019}\natexlab{}.
\newblock \showarticletitle{Pytorch lightning}.
\newblock \bibinfo{journal}{\emph{GitHub. Note: https://github.
  com/williamFalcon/pytorch-lightning Cited by}}  \bibinfo{volume}{3}
  (\bibinfo{year}{2019}).
\newblock


\bibitem[\protect\citeauthoryear{Jain, Garg, Perla, and Hebbalaguppe}{Jain
  et~al\mbox{.}}{2019}]%
        {gestarlite}
\bibfield{author}{\bibinfo{person}{Varun Jain}, \bibinfo{person}{Gaurav Garg},
  \bibinfo{person}{Ramakrishna Perla}, {and} \bibinfo{person}{Ramya
  Hebbalaguppe}.} \bibinfo{year}{2019}\natexlab{}.
\newblock \showarticletitle{GestARLite: An On-Device Pointing Finger Based
  Gestural Interface for Smartphones and Video See-Through Head-Mounts}.
\newblock \bibinfo{journal}{\emph{ArXiv}}  \bibinfo{volume}{abs/1904.09843}
  (\bibinfo{year}{2019}).
\newblock


\bibitem[\protect\citeauthoryear{Lepetit}{Lepetit}{2020}]%
        {lepetit20}
\bibfield{author}{\bibinfo{person}{Vincent Lepetit}.}
  \bibinfo{year}{2020}\natexlab{}.
\newblock \bibinfo{title}{Recent Advances in 3D Object and Hand Pose
  Estimation}.
\newblock
\newblock
\showeprint[arxiv]{2006.05927}~[cs.CV]


\bibitem[\protect\citeauthoryear{L\"{u} and Li}{L\"{u} and Li}{[n.d.]}]%
        {gcoder}
\bibfield{author}{\bibinfo{person}{Hao L\"{u}} {and} \bibinfo{person}{Yang
  Li}.} \bibinfo{year}{[n.d.]}\natexlab{}.
\newblock \showarticletitle{Gesture Coder: A Tool for Programming Multi-Touch
  Gestures by Demonstration} \emph{(\bibinfo{series}{CHI ’12})}.
\newblock


\bibitem[\protect\citeauthoryear{Lugaresi, Tang, Nash, McClanahan, Uboweja,
  Hays, Zhang, Chang, Yong, Lee, et~al\mbox{.}}{Lugaresi et~al\mbox{.}}{2019}]%
        {lugaresi2019MediaPipe}
\bibfield{author}{\bibinfo{person}{Camillo Lugaresi}, \bibinfo{person}{Jiuqiang
  Tang}, \bibinfo{person}{Hadon Nash}, \bibinfo{person}{Chris McClanahan},
  \bibinfo{person}{Esha Uboweja}, \bibinfo{person}{Michael Hays},
  \bibinfo{person}{Fan Zhang}, \bibinfo{person}{Chuo-Ling Chang},
  \bibinfo{person}{Ming~Guang Yong}, \bibinfo{person}{Juhyun Lee},
  {et~al\mbox{.}}} \bibinfo{year}{2019}\natexlab{}.
\newblock \showarticletitle{Mediapipe: A framework for building perception
  pipelines}.
\newblock \bibinfo{journal}{\emph{arXiv preprint arXiv:1906.08172}}
  (\bibinfo{year}{2019}).
\newblock


\bibitem[\protect\citeauthoryear{Oney, Krosnick, Brandt, and Myers}{Oney
  et~al\mbox{.}}{[n.d.]}]%
        {oney19}
\bibfield{author}{\bibinfo{person}{Steve Oney}, \bibinfo{person}{Rebecca
  Krosnick}, \bibinfo{person}{Joel Brandt}, {and} \bibinfo{person}{Brad
  Myers}.} \bibinfo{year}{[n.d.]}\natexlab{}.
\newblock \showarticletitle{Implementing Multi-Touch Gestures with Touch Groups
  and Cross Events}. In \bibinfo{booktitle}{\emph{CHI'19}}.
\newblock


\bibitem[\protect\citeauthoryear{Paszke}{Paszke}{[n.d.]}]%
        {pytorch}
\bibfield{author}{\bibinfo{person}{Adam et~al. Paszke}.}
  \bibinfo{year}{[n.d.]}\natexlab{}.
\newblock \showarticletitle{PyTorch: An Imperative Style, High-Performance Deep
  Learning Library}.
\newblock In \bibinfo{booktitle}{\emph{Neurips'19}}.
\newblock


\bibitem[\protect\citeauthoryear{Ren, Liu, Ding, and Liu}{Ren
  et~al\mbox{.}}{2020}]%
        {ren2020survey}
\bibfield{author}{\bibinfo{person}{Bin Ren}, \bibinfo{person}{Mengyuan Liu},
  \bibinfo{person}{Runwei Ding}, {and} \bibinfo{person}{Hong Liu}.}
  \bibinfo{year}{2020}\natexlab{}.
\newblock \bibinfo{title}{A Survey on 3D Skeleton-Based Action Recognition
  Using Learning Method}.
\newblock
\newblock
\showeprint[arxiv]{2002.05907}~[cs.CV]


\bibitem[\protect\citeauthoryear{Ruser, Vorwerg, and Eicher}{Ruser
  et~al\mbox{.}}{[n.d.]}]%
        {ruserhome20}
\bibfield{author}{\bibinfo{person}{Heinrich Ruser}, \bibinfo{person}{Susan
  Vorwerg}, {and} \bibinfo{person}{Cornelia Eicher}.}
  \bibinfo{year}{[n.d.]}\natexlab{}.
\newblock \showarticletitle{Making the Home Accessible - Experiments with an
  Infrared Handheld Gesture-Based Remote Control}. In
  \bibinfo{booktitle}{\emph{{HCI} International 2020 - Posters}}.
\newblock


\bibitem[\protect\citeauthoryear{Streeter and Gauch}{Streeter and
  Gauch}{[n.d.]}]%
        {streeter20}
\bibfield{author}{\bibinfo{person}{Lora Streeter} {and} \bibinfo{person}{John
  Gauch}.} \bibinfo{year}{[n.d.]}\natexlab{}.
\newblock \showarticletitle{Detecting Gestures Through a Gesture-Based
  Interface to Teach Introductory Programming Concepts}. In
  \bibinfo{booktitle}{\emph{{HCI}'20}},
  \bibfield{editor}{\bibinfo{person}{Masaaki Kurosu}} (Ed.).
\newblock


\bibitem[\protect\citeauthoryear{Zhang, Bazarevsky, Vakunov, Tkachenka, Sung,
  Chang, and Grundmann}{Zhang et~al\mbox{.}}{2020}]%
        {zhang2020MediaPipe}
\bibfield{author}{\bibinfo{person}{Fan Zhang}, \bibinfo{person}{Valentin
  Bazarevsky}, \bibinfo{person}{Andrey Vakunov}, \bibinfo{person}{Andrei
  Tkachenka}, \bibinfo{person}{George Sung}, \bibinfo{person}{Chuo-Ling Chang},
  {and} \bibinfo{person}{Matthias Grundmann}.} \bibinfo{year}{2020}\natexlab{}.
\newblock \showarticletitle{MediaPipe Hands: On-device Real-time Hand
  Tracking}.
\newblock \bibinfo{journal}{\emph{arXiv preprint arXiv:2006.10214}}
  (\bibinfo{year}{2020}).
\newblock


\end{thebibliography}

\end{document}